\documentclass[11pt]{article}
\usepackage{times}
\usepackage{algorithm}
\usepackage{algpseudocode}
\usepackage{algorithmicx}
\usepackage{mathtools}
\usepackage{amssymb}
\usepackage{subfigure}
\usepackage{flushend}
\begin{document}
\global\def\refname{{\normalsize \it References:}}
\baselineskip 12.5pt
%
%
% TITLE, AUTHOR, ABSTRACT, KEYWORDS
%
\title{\LARGE \bf A Genetic Algorithm for Software Design Migration from Structured to Object Oriented Paradigm}

\date{}

\author{\hspace*{-10pt}
\begin{minipage}[t]{2.3in} \normalsize \baselineskip 12.5pt
\centerline{Md. Selim}
\centerline{Institute of Information Technology}
\centerline{University of Dhaka, Bangladesh}
\centerline{selim.iitdu@gmail.com}
\end{minipage} \kern 0in
\begin{minipage}[t]{2.3in} \normalsize \baselineskip 12.5pt
\centerline{Saeed Siddik}
\centerline{Institute of Information Technology}
\centerline{University of Dhaka, Bangladesh}
\centerline{siddik.saeed@gmail.com}
\end{minipage}\kern 0in
\begin{minipage}[t]{2.3in} \normalsize \baselineskip 12.5pt
\centerline{Alim Ul Gias}
\centerline{Institute of Information Technology}
\centerline{University of Dhaka, Bangladesh}
\centerline{alimulgias@gmail.com}
\end{minipage}
\\ \\ 
\begin{minipage}[t]{3in} \normalsize \baselineskip 12.5pt
\centerline{M. Abdullah-Al-Wadud}
\centerline{Department of Industrial and Management Engineering}
\centerline{Hankuk University of Foreign Studies, South Korea}
\centerline{wadud@hufs.ac.kr}
\end{minipage} \kern 0in
\begin{minipage}[t]{3in} \normalsize \baselineskip 12.5pt
\centerline{Shah Mostafa Khaled}
\centerline{Institute of Information Technology}
\centerline{University of Dhaka, Bangladesh}
\centerline{khaled@univdhaka.edu}
\end{minipage}
%
% If you are three authors then you can use three mini--pages
% instead of two. Their horizontal size must be less than 2.7in
% indicated above. It can be e.g. 2.3in. However, you must pay
% attention that you do not exceed the total width of the text.
%
\\ \\ \hspace*{-10pt}
\begin{minipage}[b]{6.9in} \normalsize
\baselineskip 12.5pt {\it Abstract:}
% The text of the abstract follows.
The potential benefit of migrating software design from Structured to Object Oriented Paradigm is manifolded including modularity, manageability and extendability. This design migration should be automated as it will reduce the time required in manual process. Our previous work has addressed this issue in terms of optimal graph clustering problem formulated by a quadratic Integer Program (IP). However, it has been realized that solution to the IP is computationally hard and thus heuristic based methods are required to get a near optimal solution. This paper presents a Genetic Algorithm (GA) for optimal clustering with an objective of maximizing intra-cluster edges whereas minimizing the inter-cluster ones. The proposed algorithm relies on fitness based parent selection and cross-overing cluster elements to reach an optimal solution step by step. The scheme was implemented and tested against a set of real and synthetic data. The experimental results show that GA outperforms our previous works based on Greedy 
and Monte Carlo approaches by 40\% and 49.5\%. 
\\ [4mm] {\it Key--Words:}
% The key-words follow.
Software Design Migration, Optimal Graph Clustering, Genetic Algorithm
\end{minipage}
\vspace{-10pt}}

\maketitle

\thispagestyle{empty} \pagestyle{empty}
% numbers of pages are supplemented by the editor
%
% THE BEGINNING OF THE TEXT
%
\section{Introduction}
\label{intro} \vspace{-4pt}
Software design migration from Structured to Object Oriented paradigm is essential for large legacy software \cite{bennett1995legacy} due to its lack of modularity, manageability and extendability. A possible way of shifting the paradigm could be re-designing the whole product from the scratch or manual design migration which could be error-prone and time consuming. An automated Structured to Object Oriented paradigm migration could reduce those errors and time consumption hence, motivating industries to adopt the procedure.

This scenario has been represented as an optimal graph clustering problem. It has been formalized in our previous work \cite{saeed2013optimizing} with $G(V,E)$ as the underlying undirected graph of a call graph with $V$ and $E$ as the set of vertices and edges respectively, $n = \left | V \right |, m = \left | E \right |$. 

The problem of maximizing  intra-cluster edges, minimizing inter-cluster edges, and maximizing the number of clusters is used as an index to measure quality of a clustering. The matrix is referred to as Kal ($\kappa$) in the rest of this paper:
{
\footnotesize
\begin{align}
\label{kal}
\kappa = \sum_i x_i - \sum_i y_i + \sum_j \mid \mathcal{C}_j\mid
\end{align}
}
Here $x_i, y_i \in \{0,1\}$, $i=1,2,......,m$ refers to the intra-cluster edges, $x_i=1$ if $x_i$ is intra-cluster; $y_i=1$ if $y_i$ is inter-cluster. $C_j$, $j=1,2,.....,n$ represent cluster heads. $C_j \in \{0,1 \}$ if vertex $j\in V$ is the head of a cluster.

Selim \cite{selim2013thesis} proved the problem to be a computationally hard optimization problem. The optimal solution to such problems cannot be found in polynomial time, and therefore search for solution to the problem has to rely on approximation or heuristics.

This paper introduces a Genetic Algorithm for optimal graph clustering that focuses on maximizing  and minimizing the intra and inter cluster edges respectively. The fitness of each cluster is significantly depended on it's intra cluster edges. Based on the fitness of each individual clusters within a clustering, pairs are formed. A cross-over takes place within those generated pairs to exchange the vertices. Mutation within a cluster may take place based on a probability distribution.

The algorithm was implemented where a clustering scheme $C$ yielded by a greedy algorithm \cite{saeed2013optimizing} was used as an initial seed. That implementation was assessed based on 3 data instances that includes both real life and synthetically generated ones. The results show that the genetic algorithm produced better results in terms of all metrics used in \cite{saeed2013optimizing} which include clustering coefficient ($\Psi$) \cite{watts1998collective}, characteristic path length ($\chi$) \cite{watts1998collective} and Kal ($\kappa$) index.

Rest of the paper is organized as follows: Section \ref{related} reviews the research done on SP to OOP design migration, graph clustering, and presents matrices to measure clustering quality. Section \ref{method} presents the proposed genetic algorithm approach, Section \ref{result} presents the data and experimental results, Section \ref{conc} concludes the paper.

\section{Related Work}
\label{related} \vspace{-4pt}
State of the art works regarding software design migration includes automatic migration from code to design \cite{sneed1995extracting}, hierarchical clustering research in the context of software architecture recovery and modularization \cite{maqbool2007hierarchical}, architectural comparison of commercial software and scientific research software \cite{heroux2009barely} and  empirical approach for migrating from Structured Programming Code to Object Oriented Design \cite{dineh2011codetodeign}.

Franti et al. \cite{franti1997genetic} used variations of Genetic algorithm approaches for solving large scale clustering problem. They introduced three new efficient crossover techniques that are the hybrid outcome of genetic algorithm and k-means algorithm. Their proposed techniques are based on k-dimensional Euclidean distances. 

A self adaptive genetic algorithm is proposed in \cite{kivijarvi2003self} for cluster analysis that associates a set of parameters with each cluster and these parameters update by crossover and mutation. Wang et al. \cite{wang2008fuzzy} presented a fuzzy genetic algorithm for cluster analysis with c-means clustering algorithm. They used genetic algorithm for minimizing the risk of trapping in local minimum. Hruschka et al. \cite{hruschka2003genetic} proposed a genetic algorithm for finding a right number of clusters. They also used an encoding schema for determining chromosome, and Silhouette method is used for validating cluster data. Maulik et al. \cite{maulik2000genetic} done a comparative study on k-means and genetic algorithm for cluster finding. The used  n-dimensional for searching cluster centers.

Recently, Saeed et al. modeled structured to object oriented design migration as a optimal graph clustering problem which is realized as computationally hard \cite{saeed2013optimizing, selim2013thesis}. They developed certain heuristic algorithms based on Monte Carlo and Greedy approaches and formulated the Kal ($\kappa$) index for measuring the quality of a cluster. Moreover the clustering coefficient ($\Psi$) and characteristic path length ($\chi$) was used for assessing the quality.

Clustering Coefficient (CC) \cite{watts1998collective, braha2007statistical} is a measure of degree to which vertices in a graph tend to cluster together. Local clustering coefficient can be used to measure CC $(\Psi)$ index where the local clustering coefficient of a vertex quantifies how close its neighbors are to being a complete graph. Suppose, a graph $G = (V,E)$ consists of a set of vertices $V$ and a set of edges $E$. If an edge $e_{ij}$ connects vertex $v_i,v_j$, the neighborhood $N_i$ for the vertex $v_i$ is defined as its immediately connected neighbors: $N_i = \{ v_j : e_{ij} \in E \cap e_{ij} \in E \}$. Clustering Coefficient $\Psi$ of an undirected graph is defined as-
{
\footnotesize
\vspace{-.25cm}
\begin{align}
\label{cc}
\Psi = \frac{1}{N} \sum_{i=1}^{N}\Psi_i
\end{align}
}
{
\footnotesize
\vspace{-.25cm}
\begin{align*}
where \ \ \Psi_i = \frac{2\left | \{e_{ij}:v_j, v_k \in N_i, e_{jk} \in E\}\right | }{K_i(K_i-1)}
\end{align*}
}
Equation \ref{cc} $\Psi_i$ denotes the  clustering coefficient of node $i$ and $k_i$  is number of vertices connected to vertex $i$, and $n_i$ is actual number of edges within $k_i$ adjacent vertices.

Characteristics Path Length (CPL) \cite{watts1998collective, braha2007statistical} is the distance between pairs of vertices in a connected undirected graph \cite{braha2007statistical}. Let $d({v_i,v_j})$ denote the shortest distance between vertices $v_i$ and $v_j$, where $\{v_1, v_2\} \in V$ in an unweighed undirected graph $G$ . If $v_1 = v_2$ or $v_2$ cannot be reached from $v_1$ then $d({v_i,v_j}) = 0$, otherwise $d({v_i,v_j}) = 1$. Based on these definitions, Characteristics Path Length $ \chi$ of an undirected graph can defined as- 
{
\footnotesize
\vspace{-.15cm}
\begin{align}
\label{cpl}
\chi =  \frac{1}{N(N-1))}\cdot\sum_{i\neq j}d({v_i,v_j})
\end{align}
}

Review of the state of the art works show that software design migration using graph clustering did not receive high attention from the researchers. However, the scope of addressing the issue has broadened as it have been modeled in \cite{saeed2013optimizing}. Different meta heuristic based algorithms can be used utilizing the model to reach an optimal solution to the problem.

\section{Proposed Genetic Algorithm for Design Migration}
\label{method} \vspace{-4pt}

The Genetic Algorithm based meta-heuristic approach presented in this section has the underlying undirected graph $G(V,E)$ of a call graph as the input. It produces a clustering scheme $\mathcal{C}$ = $\{C_1, C_2, ... , C_k\}$ with $k$ clusters of vertices $v_i \in V$, with $\cup_{i=1..k} C_i = V$ and $\cap_{i=1..k} C_i = \emptyset$, as a clue to a modularized object oriented design. This clusters represent the underlying potential classes and/or interfaces in the future object oriented design.

\begin{algorithm}[!ht]
\caption{Genetic Algorithm for Graph Clustering}
\label{ga}
\begin{algorithmic}[1]
\footnotesize
\Require {Call Graph $G(V,E)$}
\Ensure Clustering $\mathcal{C}$, Clustering Coefficient$(\Psi)$, Characteristics Path Length $(\chi)$, Kal$(\kappa)$
\State \textbf{Begin}
\State Randomly associate a unique integer order $\tilde{o}(v)$ to all $v \in V$ so that $\cup \ \tilde{o}(v) \subset \mathbb{Z}$ and $\cap \ \tilde{o}(v)=\emptyset$
\State $\mathcal{C}_{init} \leftarrow GreedyClustering(G)$
\State $\mathcal{C} \leftarrow \mathcal{C}_{init}$
\Repeat
\State $\mathcal{C}_{tmp} \leftarrow \mathcal{C}$ 
\State $F \leftarrow FitnessCalculation(\mathcal{C}_{tmp}, G)$
\State $\mathcal{P} \leftarrow ParentSelection(\mathcal{C}_{tmp}, F)$
\State $\mathcal{C}_{tmp} \leftarrow CrossOver(\mathcal{C}_{tmp}, \mathcal{P}, \tilde{O})$
\State $\mathcal{C}_{tmp} \leftarrow Mutation(\mathcal{C}_{tmp}, \tilde{O})$
\If {$\kappa_{\mathcal{C}_{tmp}} \geq \kappa_{\mathcal{C}}$}
    \State $\mathcal{C} \leftarrow \mathcal{C}_{tmp}$
  \EndIf
\Until {$\kappa_{\mathcal{C}_{tmp}}$ does not improve for $t$ consecutive iterations}
\State Calculate $\Psi$, $\chi$, $\kappa$ using $\mathcal{C}$ and Eq. (\ref{cc}), (\ref{cpl}) and (\ref{kal})
\State \textbf{End}
\end{algorithmic}
\end{algorithm}

\begin{algorithm}[!ht]
\caption{GreedyClustering \cite{saeed2013optimizing}}
\label{greedy}
\begin{algorithmic}[1]
\footnotesize
\Require {Call Graph $G(V,E)$}
\Ensure Clustering $\mathcal{C}$
\State \textbf{Begin}
\State Fix initial number of clusters $\mathcal{C}_{i=1,...,n}$ to $n=\sqrt{\left |  V \right |}$
\State Pick unique vertex $v_i \in V$ in decreasing order of vertex degree and a make one-to-one correspondence assignment of $v_i$ to $\mathcal{C}_j$, where $i,j = 1,2,3 ... n$
\ForAll{edge $e \in E$}
 \State Assume $v_1$ and $v_2$ be the two end points of $e$
  \If {$v_1 \in \mathcal{C}_i$ and $v_2$ unassigned to any cluster}
    \State $\mathcal{C}_i \leftarrow \mathcal{C}_i \cup \{v_2\}$
  \ElsIf {$v_1$ unassigned to any cluster and $v_2 \in \mathcal{C}_j$}
    \State $\mathcal{C}_j \leftarrow \mathcal{C}_j \cup \{v_1\}$
  \EndIf
\EndFor
\State \textbf{End}
\end{algorithmic}
\end{algorithm}

The proposed scheme for optimal graph clustering using genetic algorithm is presented in Algorithm \ref{ga}. The initial seed for the algorithm, a clustering $\mathcal{C}_{init}$ is generated by Algorithm \ref{greedy}, a greedy heuristics reported in our previous work \cite{saeed2013optimizing}. The solution $\mathcal{C}_{init}$ is considered as the first candidate solution. This candidate solution is iteratively modified using the operations of genetic algorithm meta-heuristic in the search for a better solution. This algorithm stops when the current best solution cannot be further improved for a number of $t$ consecutive steps. The algorithm performs the following tasks iteratively:

\begin{enumerate}
 \item Measure the fitness $(f_i)$ of each cluster $(C_i)$ for all clusters in the solution
 \item Create cluster pairs $(p)$ based on the fitness
 \item Perform cross-over between the clusters $C_a$ and $C_b$ in a pair $p_i = (a,b)$ by exchanging member vertices of the clusters, for all pairs $p \in P$ 
 \item Perform mutation by changing the order of vertices within a randomly picked cluster
 \item Compare the $\kappa$ of the candidate solution in hand with the best solution found so far. If $\kappa$ for candidate solution is better, change the best solution to the candidate solution
\end{enumerate}

\begin{algorithm}[!ht]
\caption{FitnessCalculation}
\label{cf}
\begin{algorithmic}[1]
\footnotesize
\Require {Clustering $\mathcal{C}=\{C_1, C_2, ...C_m\}$, Call Graph $G(V,E)$}
\Ensure List of fitness $F={f_1, f_2, ...f_m}$ where $f_i$ is the fitness of cluster $C_i$
\State \textbf{Begin}
\ForAll{ $C_i \in \mathcal{C}$}
\State $n \leftarrow 0$
  \ForAll {pair $v_j, v_u \in C_i$}
  \If {$(v_j,v_u) \in E$}
  \State $n \leftarrow n+1$
  \EndIf
  \EndFor
 \State Compute $\chi_{c_i}$ using Eq. \ref{cpl}
 \State  $f_i \leftarrow n+\chi_{c_i}$
\EndFor
\State \textbf{End}
\end{algorithmic}
\end{algorithm}

\begin{algorithm}[!ht]
\caption{ParentSelection}
\label{ps}
\begin{algorithmic}[1]
\footnotesize
\Require {Clustering $\mathcal{C}=\{C_1, C_2, ...C_m\}$, Fitness list $F={f_1, f_2, ...f_m}$}
\Ensure List of cluster index pairs $\mathcal{P}=\{p_1, p_2, ...p_{\lceil{\frac{m}{2}}\rceil}\}$ where $p_i=\{\mathbb{N},\mathbb{N}\}$
\State \textbf{Begin}
\State Order $C_i \in \mathcal{C}$ in order of $f_i$ associated with $C_i \ \ \ \forall_{i=1,2,...m}$ to produce list $\mathcal{C}_{ordered}$
\State List $\mathcal{P} \leftarrow \emptyset$
\State $i \leftarrow 0$
\ForAll{ $C_i, C_{i+1} \in \mathcal{C}_{ordered}$\ , $i=1, 2, ...|\mathcal{C}_{ordered}|-1$ in order of $\mathcal{C}_{ordered}$}
\State add $(i,i+1)$ to list $\mathcal{P}$
\State $i \leftarrow i+2$
\EndFor
\State \textbf{End}
\end{algorithmic}
\end{algorithm}

\begin{algorithm}
\caption{CrossOver}
\label{co}
\begin{algorithmic}[1]
\footnotesize
\Require {Clustering $\mathcal{C}$, List of cluster index pairs $\mathcal{P}=\{p_1, p_2, ...p_{\lceil{\frac{m}{2}}\rceil}\}$, order $\tilde{o}$ of $v \in V$}
\Ensure Clustering $\mathcal{C}$
\State \textbf{Begin}
\ForAll{ $p_i \in \mathcal{P}$}
\State $\mathcal{C}^\prime \leftarrow \mathcal{C}$
\State Generate random number $r_1$ and $r_2$ such that $1 < r_1 < |\mathcal{C}_a|$\ , $1 < r_2 < |\mathcal{C}_b|$\ , $\{a,b\} \in p_i$ 
\State $\mathcal{C}_{a_{tmp}} \leftarrow \mathcal{C}_a$
\State $\mathcal{C}_{b_{tmp}} \leftarrow \mathcal{C}_b$
\For {$i \in 1:r_1$}
\State Pick $v_i \in \mathcal{C}_{a_{tmp}}$ in desc. order of $\tilde{o}(v_i) \in \mathcal{C}_{a_{tmp}}$
\State $\alpha \leftarrow \alpha \cup v_i$
\State $\mathcal{C}_{a_{tmp}} \leftarrow \mathcal{C}_{a_{tmp}} \setminus v_i$
\EndFor
\For {$j \in 1:r_2$}
\State Pick $v_j \in \mathcal{C}_{b_{tmp}}$ in desc. order of $\tilde{o}(v_j) \in \mathcal{C}_{b_{tmp}}$
\State $\beta \leftarrow \beta \cup v_j$
\State $\mathcal{C}_{b_{tmp}} \leftarrow \mathcal{C}_{b_{tmp}} \setminus v_j$
\EndFor
\State $\mathcal{C}_a^\prime \leftarrow \mathcal{C}_{a_{tmp}} \cup \beta$
\State $\mathcal{C}_b^\prime \leftarrow \mathcal{C}_{b_{tmp}} \cup \alpha$
\State $\mathcal{C}^\prime \leftarrow \mathcal{C} \setminus \mathcal{C}_a$
\State $\mathcal{C}^\prime \leftarrow \mathcal{C} \setminus \mathcal{C}_b$
\State $\mathcal{C}^\prime \leftarrow \mathcal{C} \cup \mathcal{C}_a^\prime \cup \mathcal{C}_b^\prime$
\State $\mathcal{C}^\prime \leftarrow \mathcal{C} \cup \mathcal{C}_b^\prime$
\EndFor
\State $\mathcal{C} \leftarrow \mathcal{C}^\prime$
\State \textbf{End}
\end{algorithmic}
\end{algorithm}

\begin{algorithm}
\caption{Mutation}
\label{mut}
\begin{algorithmic}[1]
\footnotesize
\Require {Clustering $\mathcal{C}$, Order $\tilde{O}$ of $v \in V$}
\Ensure Clustering $\mathcal{C}$
\State \textbf{Begin}
\State $\mathcal{C}^\prime \leftarrow \mathcal{C}$
\State Generate a random number $r \in \mathbb{R}$ between 0 to 1
\If {$r \leq \epsilon$}
\State Randomly pick  $C \in \mathcal{C}$
\State Randomly pick $v_a, v_b \in c$
\State ${tmp} \leftarrow \tilde{o}(v_a)$
\State $\tilde{O}(v_a) \leftarrow \tilde{o}(v_b)$
\State $\tilde{o}(v_b) \leftarrow {tmp}$
\EndIf
\State $\mathcal{C} \leftarrow \mathcal{C}^\prime$
\State \textbf{End}
\end{algorithmic}
\end{algorithm}

The steps involving fitness calculation and parent selection is illustrated in Algorithm \ref{cf} and \ref{ps}. Two measures have been used for fitness calculation that includes the number of intra-cluster edge and characteristic path length ($\chi$). The fitness of each cluster is realized as the summation of those two measures. The result of this step is a fitness list $(F)$ which works as the basis of Parent Selection. This step involves sorting the fitness list in descending order and pairing the clusters, corresponding to a fitness value, from top to bottom. It produces a list of cluster pairs $P=\{p_1, .. ,p_{\lfloor m/2\rfloor}\}$ for cross-over.

Algorithm \ref{co} and \ref{mut} illustrates the procedure involving cross-over and mutation. A random number $\tilde{o}(v)$ is assigned to each vertex $v \in V$, which is used in mutation and cross-over operations. The cross-over operation uses this order to select vertices to be exchanged between clusters. The mutation operation exchanges the order of two vertices within a cluster. Cross-over will take place for each pair that was generated during Parent Selection. After completing the cross-over, mutation takes place. However, this will occur based on a probability distribution. The process involves shuffling the order of two vertices from a randomly picked cluster.

\section{Experimental Results}
\label{result} \vspace{-4pt}

Our proposed genetic algorithm has been implemented using C++ programming language on a 32-bit Ubuntu 12.04 Operating System, 2.1 GHz Dual Core processor, 1 GB RAM computer.

\begin{table}
\label{table_ds}
\footnotesize
\centering
\caption{Number of Vertices and Edges of experimental dataset}
\begin{tabular}{l|l|l}
\hline \hline
Dataset  &	Number of Vertices&	Number of edges  \\	\hline	
BTF	 &	14	&	31	\\	
RBIo	 &	61	&	372	\\	
Synthetic166&	166	&	450	\\
\hline\hline
\end{tabular}
\end{table}

Three different datasets used in \cite{saeed2013optimizing} have been used to experiment with our proposed genetic algorithm. \emph{BTF}, \emph{RBIo} were generated from two different scientific software and \emph{Synthetic166} was synthetically generated. Table 1 describes the data set in terms of the number of user defined functions and function calls.

\begin{table}
\label{num_cluster}
\footnotesize
\centering
\caption{Number of clusters produced by proposed heuristics}
\begin{tabular}{l|l|l|l}
\hline \hline
Dataset  &	Monte Carlo&	Greedy   & Genetic  \\	\hline	
BTF	 &	3	&	4	&	4	\\	
RBIo	 &	7	&	8	&	8	\\	
Synthetic166&	23	&	13	&	13	\\
\hline\hline
\end{tabular}
\end{table}

Table 2 presents the number of clusters generated by our proposed algorithm in contrast with the algorithms in \cite{saeed2013optimizing}. Our proposed algorithm does not change the number of clusters from the initial seed, it just enhances the solution quality. Using the proposed algorithm  4, 8, and 13 clusters were obtained for dataset \emph{BTF}, \emph{RBIo}, and \emph{Synthetic166} respectively.

\begin{table}
\label{table_result} 
\footnotesize
\centering
\caption{Performance of Genetic Algorithm on Different Datasets}
\begin{tabular}{l|l|l|l|l|l|l}
\hline \hline
Dataset  &	\multicolumn{2}{ |c| }{CC($\Psi$)}&	\multicolumn{2}{ |c| }{CPL($\chi$)}&	\multicolumn{2}{|c}{Kal($\kappa$)}	\\	\hline
	 &	Seed	 	&	Final	&	Seed	 &	Final		&	Seed	 &	Final	\\	\hline
BTF	 &	0.308333	&	0.315972&	1.20536 &	0.794444	&	5	&	9	\\	
RBIo	 &	0.474293	&	0.521471&	1.52528	&	0.378843	&	5	&	77	\\	\hline
Synthetic166&	0.333908	&	0.390482&	4.10824	&	3.75334		&	35	&	135	\\
\hline\hline
\end{tabular}
\end{table}

\begin{figure}
\label{fig}
\centering
\subfigure[Clustering Coefficient$\left ( \Psi \right )$]{
  \label{cco-line}
  \includegraphics[scale =0.42]{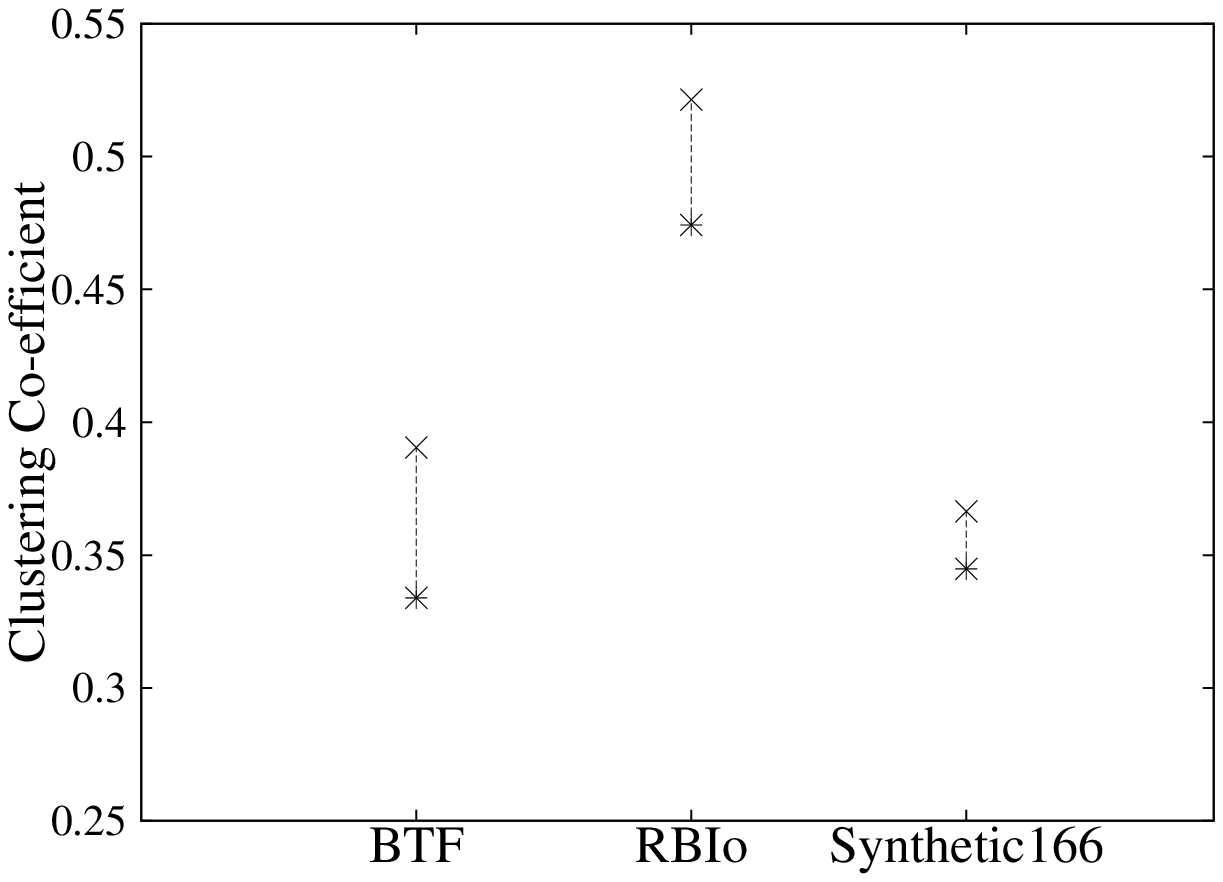}
}
\subfigure[Characteristics Path Length$\left ( \chi \right )$]{
  \label{cpl_line}
  \includegraphics[scale =0.42]{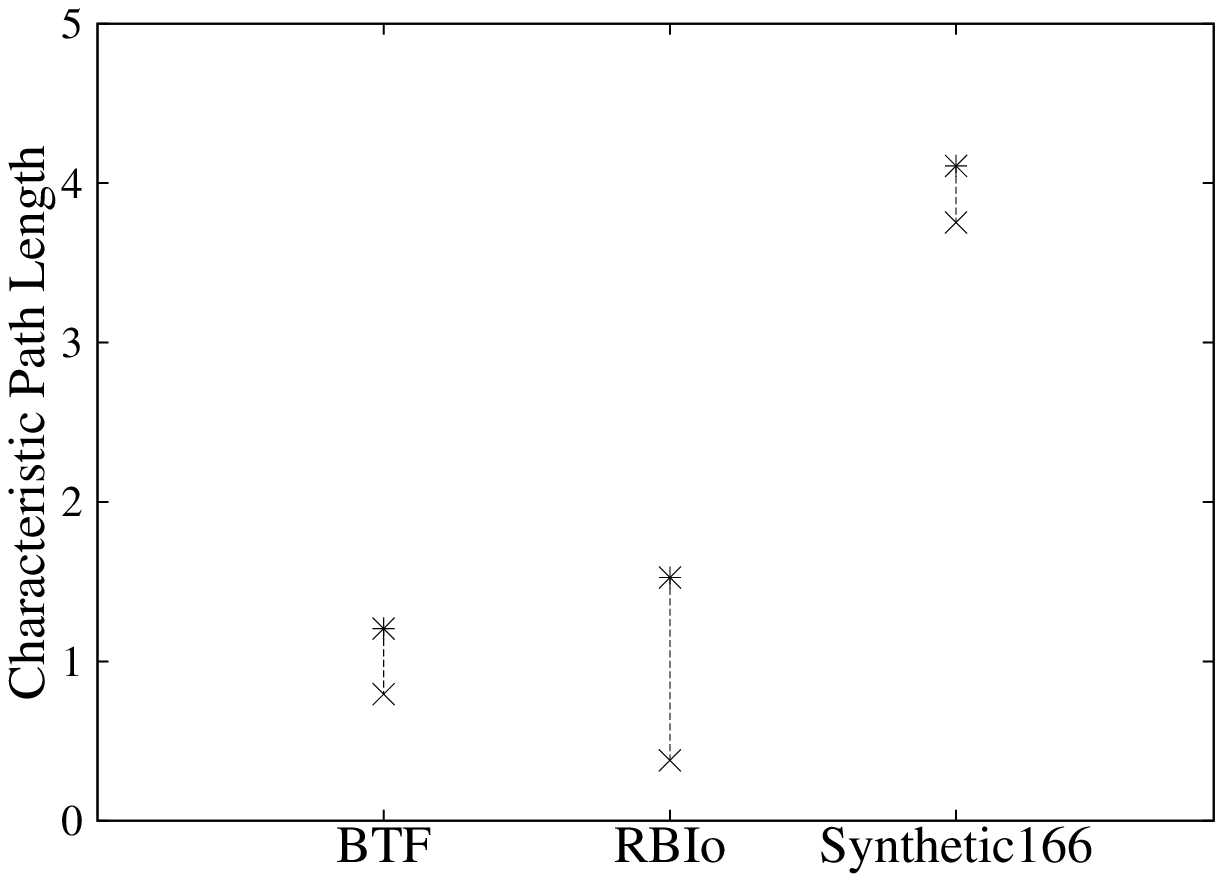}
}
\subfigure[KAL index $\left ( \kappa \right )$]{
  \label{kal_line}
  \includegraphics[scale =0.42]{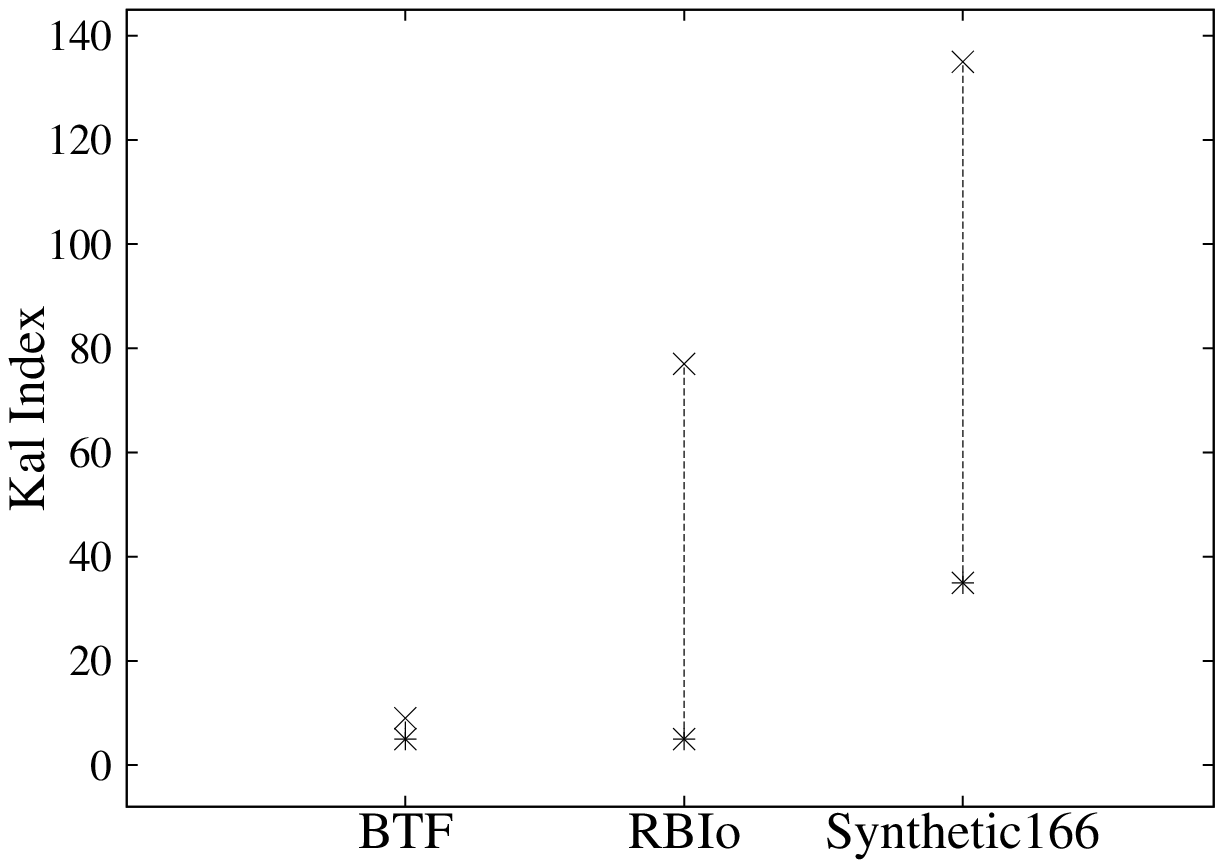}
}
\caption{Genetic Algorithm on \emph{BTF}, \emph{RBIo} and \emph{synthetic166}}

\end{figure}

Results obtained by applying Algorithm \ref{ga} on the datasets are presented in Table 3. The scores for CC($\Psi$), CPL($\chi$) and Kal($\kappa$) have been improved by 0.097, 0.39 and 5.66 times respectively using our proposed algorithm. Figure 1 presents a graphical representation of data in Table 3, here star (\textasteriskcentered) symbol refers to the score of the initial seed and cross ($\times$) symbol represent result obtained by our proposed algorithm.

\begin{figure}[h]
\label{fi} 
\centering
\includegraphics[scale =0.6]{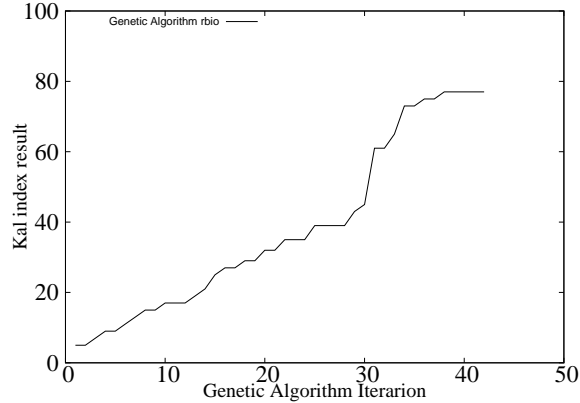}
\caption{Iteration of Genetic Algorithms on \emph{RBIo}}
\end{figure}

Figure 2 presents the gradual improvement done by our proposed algorithm on dataset \emph{RBIo}.The $\kappa$-index score of initial seed was 5, it improved over the next 43 iterations of the algorithm to a score of 77. This score, since did not improve over the next 5 iterations, has been reported as the best solution obtained.

\begin{figure}[h]
\label{fc} 
\centering
  \includegraphics[scale =0.6]{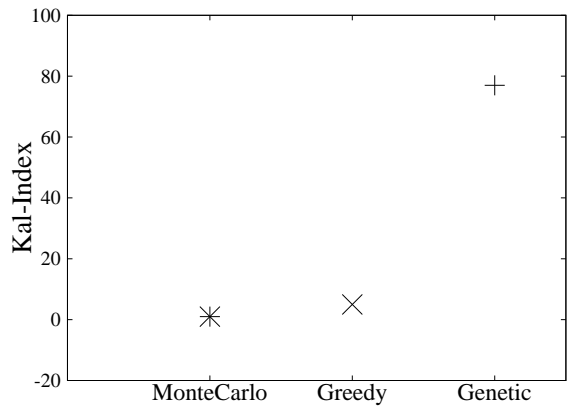}
\caption{Comparison of Kal($\kappa$) on \emph{RBIo}}
\end{figure}

Figure 3 presents the comparison of average scores of Monte Carlo based algorithms and scores of Greedy approaches \cite{saeed2013optimizing} with the scores obtained by proposed algorithm results. Sign ($*$), ($x$), and ($+$) denotes the Kal($\kappa$) index of Monte Carlo, Greedy, and Genetic Algorithm of dataset \emph{RBIo}. This figure indicates that our proposed genetic algorithm produces significantly better result than the Monte Carlo and greedy algorithms.  

\section{Conclusion}
\label{conc} \vspace{-4pt}
This paper addressed a design migration problem from Structured Language to Object Orient Paradigm. Here, we proposed a Genetic Algorithm based meta-heuristic approach and presented the test result on datasets reported in \cite{saeed2013optimizing}. Our proposed approach achieved 40\% improvement compared to greedy algorithms and 49.5\% improvement compared to the Monte Carlo approaches presented in \cite{saeed2013optimizing}.

In future we are interested to enhance the performance of the proposed algorithm trying variations of the \emph{FitnessCalculation} and \emph{ParentSelection} functions. The \emph{ParentSelection} function in our proposed algorithm selects pairs of clusters in order of fitness. Thus two high fitness clusters are crossed over. We identify this as a potential area of improvement, where we want to cross over the low fitness clusters with the high ones to see all clusters have a high fitness. The \emph{FitnessCalculation} function may be enhanced inspired by the $\kappa$ index, which to our understanding, is the most suitable matrix to measure the strength of a clustering scheme.

Currently, our research group is working towards developing a local search based algorithm to find an approximate solution to the problem. We are also interested in developing an Ant Colony Optimization based meta-heuristic approach for the problem. It would be great to be able to validate OOP design clue generated by the algorithms by practicing OOP professionals.

\vspace{10pt} \noindent
{\bf Acknowledgements:} \ This research was conducted by Optimization Research Group of Institute of Information Technology, University of Dhaka. 

Our sincere gratitude to Dr. Shahadat Hossain, Associate Professor, Dept. Math \& Computer Science,  Univ. of Lethbridge, AB, Canada for presenting this problem to us and thanks to Mr. Ahmed Tahsin Zulkernaine for providing sample datasets.

\end{document}